\documentclass[twocolumn,printnumbers,amsmath,amssymb,10pt]{revtex4}

\newcommand{\be}{\begin{equation}}
\newcommand{\ee}{\end{equation}}
\newcommand{\ba}{\begin{eqnarray}}
\newcommand{\ea}{\end{eqnarray}}
\usepackage{graphicx}
\usepackage{dcolumn}
\usepackage{bm}
\usepackage{setspace}
\usepackage{color}

\begin{document}


\title{Bivalent defect configurations in inhomogeneous nematic shells}

\author{Vinzenz Koning $^{\dag}$, Teresa Lopez-Leon $^{\dag\dag}$, Alberto Fernandez-Nieves  $^{*}$ and V. Vitelli$^{\dag}$}
\affiliation{$^{\dag}$ Instituut-Lorentz for Theoretical Physics, Universiteit Leiden, 2300 RA Leiden, The Netherlands \\
$^{\dag \dag}$ Universite Montpellier 2, Laboratoire Charles Coulomb UMR 5221, F-34095 \\
$^{*}$ ~School of Physics, Georgia Institute of Technology, Atlanta, Georgia 30332, USA. 
}

\begin{abstract} 
\noindent 
We present a theoretical study of the director fields and energetics of nematic liquid crystal shells with two pairs of surface defects. The pairs of defects can undergo abrupt transitions between a configuration of maximum separation to at state in which the defects are confined to the thinnest hemisphere. We construct a phase diagram that maps out the stability and coexistence of these two configurations as a function of shell thickness and thickness inhomogeneity. Our results compare favorably with the experimentally observed transitions in nematic double emulsion droplets and explain their hysteretic character.
\end{abstract}

\maketitle

\section{Introduction}
Many systems in condensed matter physics and elasticity can be treated
as two-dimensional, though only very few, like graphene \cite{2004Sci...306..666N} and colloidal
crystals at liquid-liquid interfaces \cite{2002Sci...298.1006D,2003Sci...299.1716B,2010Natur.468..947I}, are truly \textit{mono}layers. The theory of plates
and membranes \cite{Landau}, superfluid \cite{RevModPhys.82.1301} and liquid crystal films \cite{2009AdPhy..58..449B} can all be neatly
described by a reduction of the number of spatial dimensions from
three to two by assuming that the thickness is small compared to the
other two dimensions and approximately constant.  This reduction of
dimensions usually simplifies the analysis significantly, because the
number of variables to solve for is reduced and in addition one can employ well developed mathematical machinery
such as complex analysis that is well suited to tackle two-dimensional
problems. For instance, the use of conformal mappings has been applied succesfully in
superfluid films \cite{RevModPhys.82.1301}. Another example of the use of a conformal mapping was in
the study of orientational order on a spherical surface
\cite{1992JPhy2...2..371L}. It was found that in the ground state four defects of
charge one-half reside on the vertices of a tetrahedron, inspiring the idea of self-assembly of liquid crystal coated
particles into a diamond structure \cite{2002NanoL...2.1125N}, which in turn triggered a tremendous research activity \cite{2006PhRvE..74b1711V,2007PhRvL..99o7801F,PhysRevLett.100.197802,2008SMat....4.2059B,2008JChPh.128j4707B,2008PhRvL.101c7802S,B917180K,2011NatPh...7..391L,2011PhRvL.106x7802L,2011PhRvL.106x7801L,Lopez-Leon:2011fk,C0SM00378F,PhysRevE.85.061701,PhysRevLett.108.207803,PhysRevLett.108.057801,PhysRevE.85.061710,2012PhRvE..86b0705S,PhysRevE.86.020703,Napoli201366}. Since the defects form very
distinct regions on sphere, they can be functionalised chemically,
an idea that has been realised for a similar
system, namely metal
nanoparticles coated with a monolayer of tilted
molecules \cite{2007Sci...315..358D}. Experimental model systems of spherical nematics have
revealed that the thickness plays a crucial role \cite{2011NatPh...7..391L}. Besides the
tetravalent ground state which contains charge one-half defects lines,
bivalent defect configurations with charge-one defects have also been
observed. The reason for this is the finite thickness of the liquid
crystal shell. The charge one defect lines can lower their energies by
escaping in the third dimension, thereby leaving two pairs of point defects at the interfaces, as an alternative to splitting
into two charge one-half defects. Furthermore, the nematic liquid crystal shells generated by the
encapsulation of a water droplet by a nematic droplet are strongly
inhomogeneous in thickness due to buoyancy. Similar to effects on the buckling and folding of inhomogeneous solid capsules \cite{2012PhRvL.109m4302D}, the inhomogeneity has pronounced effects on the mechanics of the liquid crystal. 
One of the striking manifestations of the inhomogeneity 
are very abrupt confinement and deconfinement transitions in shells with two pairs of surface defects. The investigation of
this phenomenon, and more generally the theoretical study of the director fields and energetics of inhomogeneous bivalent nematic shells, is the main concern of this article. Although the thickness truly makes this a three-dimensional
problem, we are able to use two-dimensional techniques such as conformal
mappings to find an Ansatz for the director field in spherical
shells. This method is presented in detail in
section \ref{sec:calculation}.  In sections \ref{sec:homogeneous}, we study the homogeneous shells
as a function of thickness, taking
into account the elastic anisotropies. In section
\ref{sec:inhomogeneous}, we construct a phase diagram for inhomogeneous shells that maps out the stability and coexistence of the confined and deconfined configurations as a function of shell thickness and thickness inhomogeneity. Our findings are shown to be in qualitative agreement with recent experimental studies. Finally, the effect of elastic
anisotropy on the deconfinement transition is briefly discussed in the concluding section \ref{sec:conclusion}.

\section{Director fields in bivalent nematic shells}
\label{sec:calculation}
\begin{figure}[h]
\centering
\includegraphics[width=0.5\columnwidth]{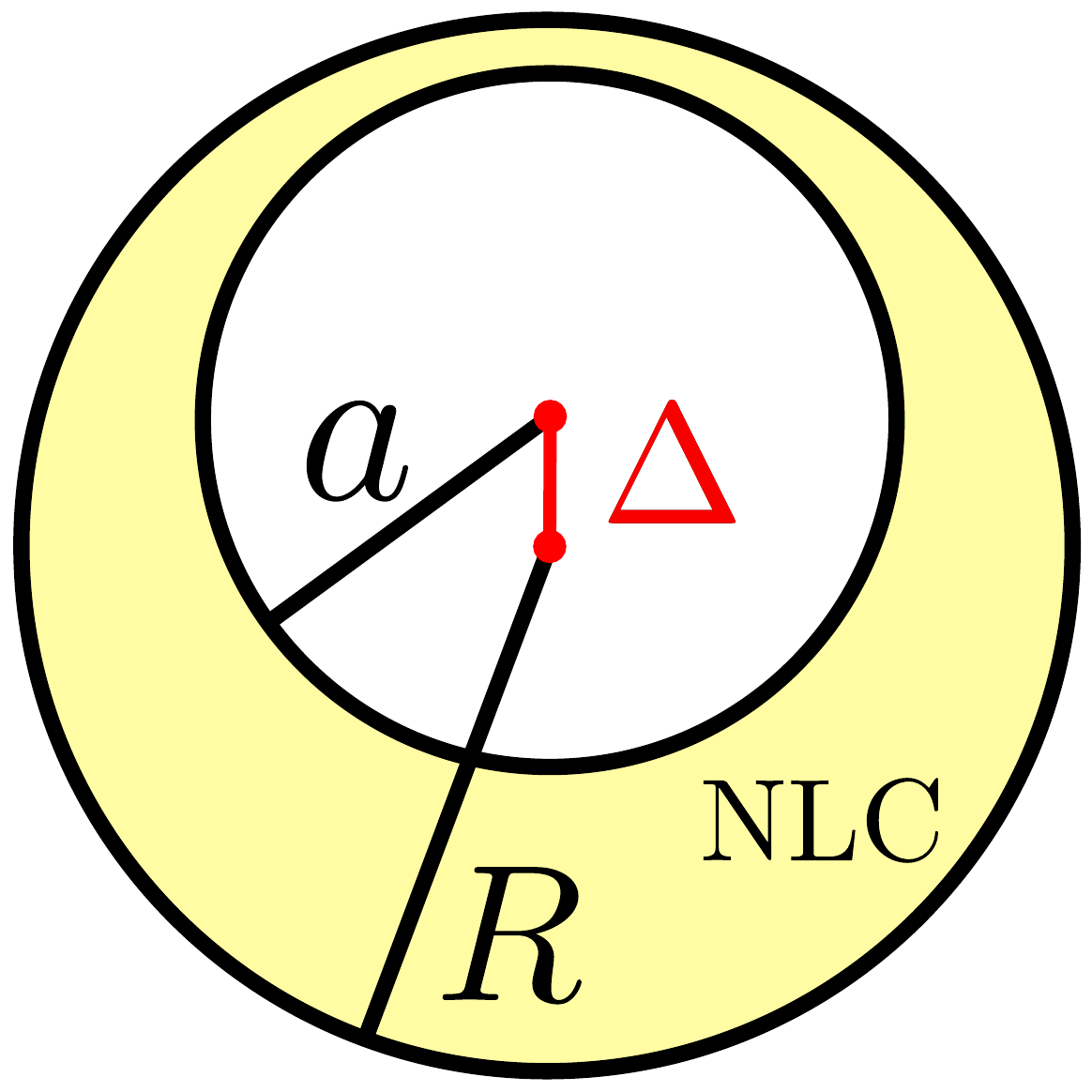}
\caption{Schematic of a nematic double emulsion droplet of radius $R$. The inner water droplet of radius $a$ is displaced by an amount $\Delta$ along the vertical direction, thereby making the top of the shell thinner. }
\label{fig:droplet}
\end{figure}

The experimental system under consideration is a nematic double emulsion
droplet: a nematic liquid crystal droplet of radius $R$ that encapsulates a smaller
water droplet of radius $a$, as depicted in Figure \ref{fig:droplet}.
In this way, we can create spherical shells of nematic
liquid crystal. We define a thickness $h \equiv R-a$ of the
shell. Since in general the  displacement of the inner water droplet
out of the centre of the nematic droplet, $\Delta$, is nonzero, $h$
should be thought of as an average quantity. A surfactant or polymer is added to the inner and outer water phases for two reasons. First of all, it stabilises the double emulsion
droplet, because it prevents the inner water droplet to coalesce with
the continuous water phase. Secondly, it anchors the nematic molecules
parallel to the interfaces. 
In modeling this experimental system we will employ elasticity theory
for nematic liquid crystals, in which one constructs a  Frank free
energy functional as an expansion in spatial distortions of the local
average orientation of the molecules, i.e. the unit director field, $\mathbf{n} \left( \mathbf{x} \right)$, that respect
the symmetries of the nematic liquid crystal \cite{deGennes,SoftMatter}:
\begin{equation}
\begin{split}
&F[\mathbf{n}\left(\mathbf{x}\right)] = \frac{1}{2} \int dV \left( K_1 \left( \nabla \cdot \bf{n} \right)^2
+ K_2 \left( \mathbf{n} \cdot \nabla \times \mathbf{n} \right)^2 \right. \\ 
&+ \left. K_3 \left( \mathbf{n} \times \nabla \times \mathbf{n} \right)^2 \right)
- K_{24} \int \mathbf{dS} \cdot \left( \mathbf{n} \nabla \cdot
  \mathbf{n} +   \mathbf{n} \times \nabla \times \mathbf{n} \right)
  \end{split}
\end{equation}
provided that we assume that these deformations are small on the
molecular lengthscale. Here, $K_1$, $K_2$, $K_3$ and $K_{24}$ are
elastic constants measuring the amount of splay, twist, bend and
saddle-splay deformations respectively.
In most of the work presented, we will work in the one-constant approximation, in which
the splay, twist and bend elastic constant are taken to be equal:
$K=K_1=K_2=K_3$. Furthermore, we discard the surface term, effectively
taking $2K_{24}=K$.
For a typical droplet size of $50\:\mu m$ the anchoring energy is much
larger than this elastic energy. Therefore, we
can take the preferred tangential alignment of the nematic molecules at
the interface as a constraint, thus forming a boundary condition
complementing the free energy.  
Our approach to minimising the free energy with respect to the
director field, will be to find a realistic Ansatz given
certain locations of the defects. By varying these locations for different shell geometries we obtain the energy landscape as a function of defect
positions, thickness and thickness inhomogeneity. The technique we employ
to obtain the Ansatz is the method of comformal mappings.
With the inverse stereographic projection we can find an Ansatz for a
director field in a homogeneous shell (section \ref{subsec:stereo}). Then, by using the electrostatic analogy
we can expand the Ansatz to the inhomogeneous case (section \ref{subsec:electrostatic_analogy}). An additional
numerical minimisation takes care of the escape of the disclination
lines in the third dimension. 

\subsection{The inverse sterographic projection and the Ansatz for the
  homogeneous shell}
\label{subsec:stereo}
\begin{figure}[h]
\centering
\includegraphics[width=\columnwidth]{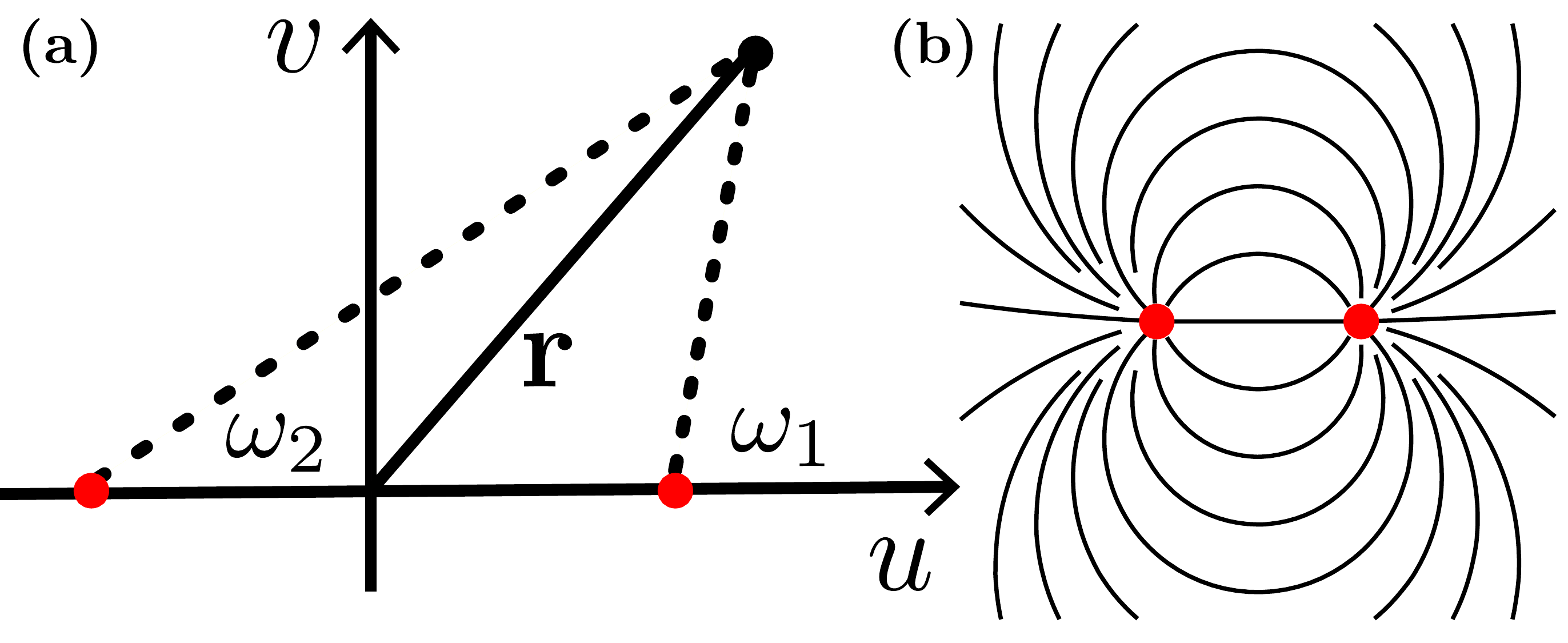}
\caption{(a) Point defects (red dots) in the $uv$-plane located at $\left( u_i, 0 \right)$. $\Phi$ at $\mathbf{ r }$ is the sum of the single defect solutions $\Phi_i = \omega_i$. (b) Schematic of the resulting fieldlines.}
\label{fig:calc}
\end{figure}

\begin{figure}[h] 
\centering
\includegraphics[width=0.6\columnwidth]{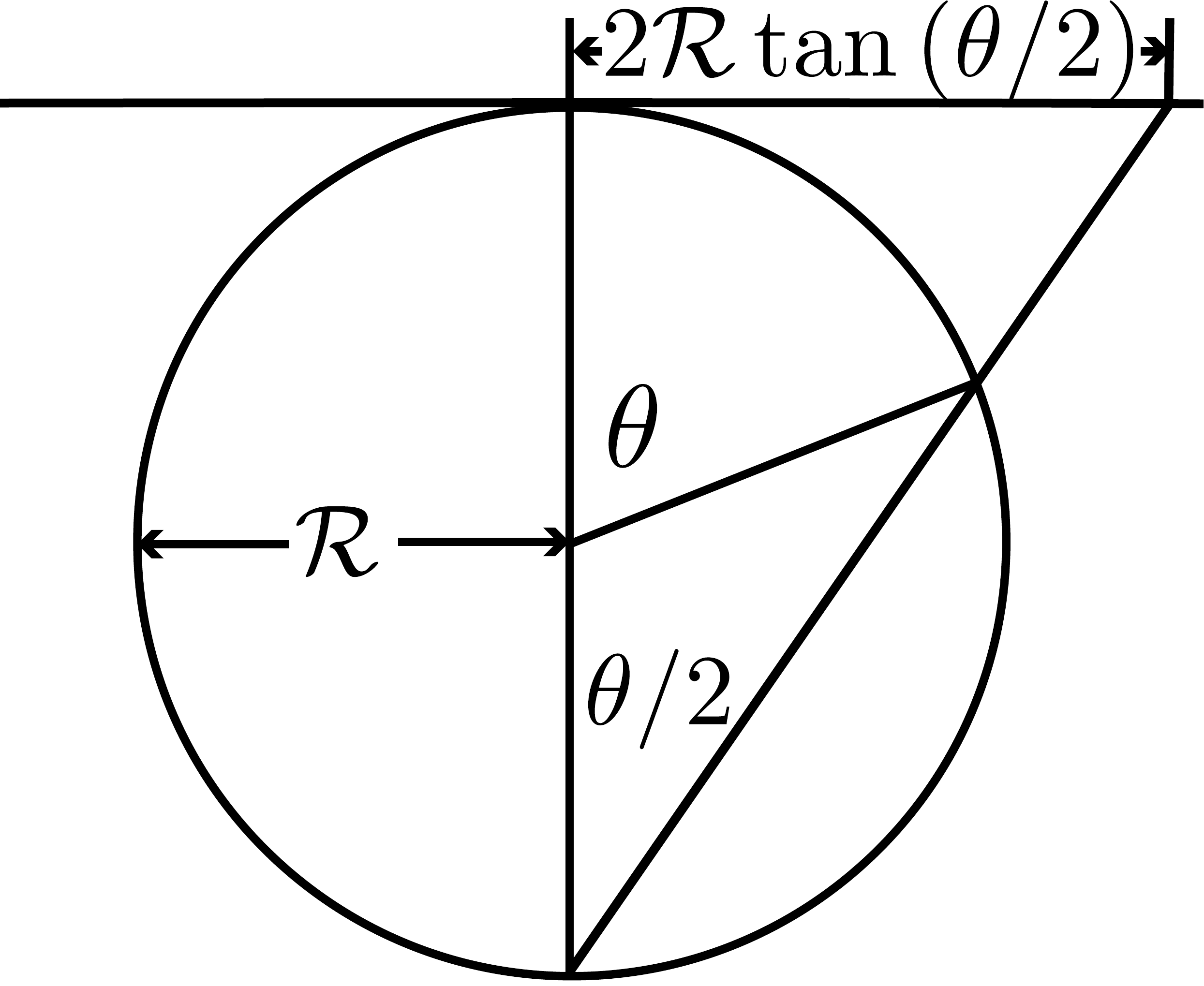}
\caption{The inverse stereographic mapping given by eq. \eqref{eq:stereo} of the the $uv$-plane onto the sphere with radius $\mathcal{R}$.}
\label{fig:stereo}
\end{figure}

\begin{figure}[h]
\centering
\includegraphics[width=0.5\columnwidth]{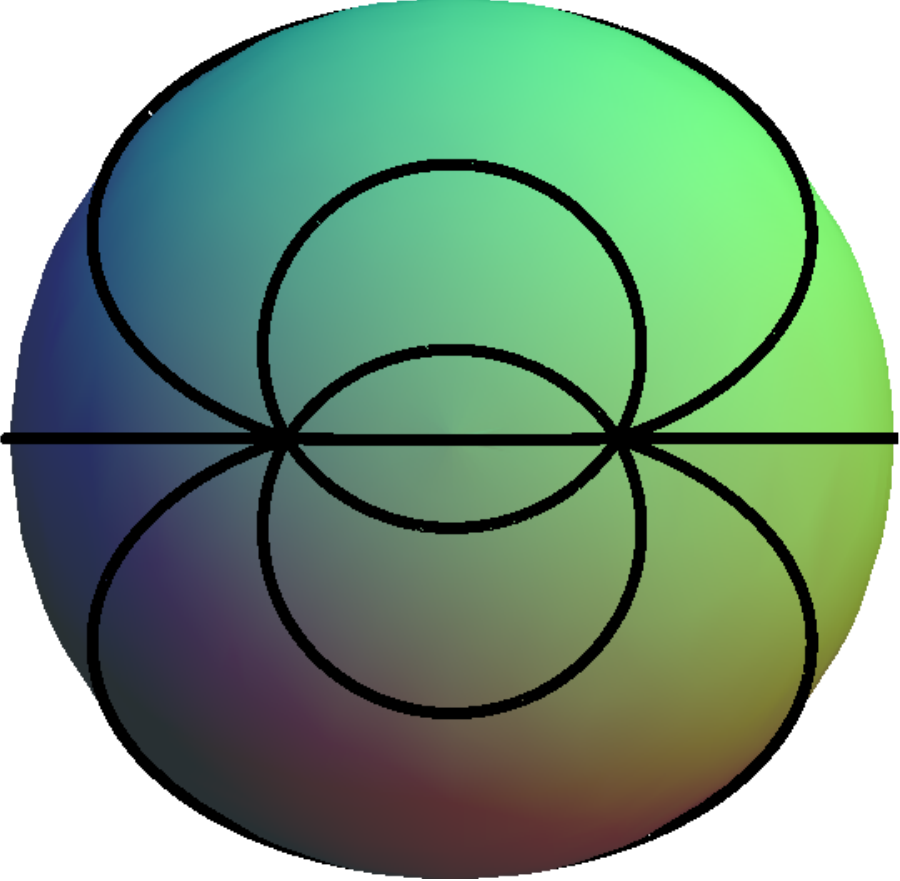}
\caption{Top view of the director field on the sphere, given by eq. (\ref{eq:ansatz_homogeneous}).}
\label{fig:splitted sphere}
\end{figure}

The Ansatz for the director $\bf{n}$ of the homogeneous bipolar shell, with two
straight disclination lines along the $z$-axis, simply reads
\begin{equation}
\label{eq:ansatz_homogeneous_bipolar}
\mathbf{n}\left(\mathbf{x}\right)=\cos{\alpha}~\hat{\theta} + \sin{\alpha}~\hat \phi,
\end{equation}
where $\hat{\theta}$ and $\hat{\phi}$ are the unit vectors corresponding to the
zenith, $\theta$, and azimuthal, $\phi$, angles
respectively. Note that $\alpha$ is the angle over which $\mathbf{n}=\hat{\theta}$ is rotated at each point on the sphere with respect to a orthonormal reference frame. Thus, the director fieldlines for $\alpha=0$ and $\alpha=\pi / 2$ correspond to the meridians and circles of latitude.
 To find the Ansatz for any other locations of the the disclination
 lines, however, we perform an inverse stereographic projection.  A
 director field in the flat plane, minimising the free energy, is
 projected onto the concentric surfaces of spheres with radii, $\mathcal{R}$, varying
 between $a$ and $R$, i.e. $a\leq \mathcal{R} \leq R $, that fill up the
 shell. Hereby, angles are preserved, i.e. this mapping is
 conformal. This director field contains two charge-one point
 defects, as we wish to eventually construct an Ansatz with two charge one line
 defects spanning the shell. Since the Euler-Lagrange equation, valid
 everywhere except at the defect cores,  
 for the angular field, $\Phi$, defined as
 the angle of the director with a cartesian reference frame $(u,v)$, is
 Laplace's equation, it obeys the superposition principle.
Therefore, the director field can be written as the sum of
the director fields of two individual defects, $\Phi_{ 1 }$ and $\Phi_{ 2 }$, positioned at $\mathbf{r}_1$ and
$\mathbf{r}_2$. Thus we write, 
\begin{equation}
\Phi \left( \mathbf{r} \right) = \Phi_1 + \Phi_2 = \alpha + \omega_{1} + \omega_{2},
\end{equation}
where $\omega_{i}$ is the azimuthal angle in the reference frame that has
$\mathbf{r}_i$ as its origin, as is shown in Fig.
\ref{fig:calc}a. Here, $\alpha$ is again a global constant.
If we take $\mathbf{r}_i = \left( u_i, 0 \right)$ the defects lie on the $u$-axis.
We have
\begin{equation}
\omega_i = \arctan \left( v, u-u_i \right), 
\end{equation}
where the two-argument $\arctan \left( y, x \right)$ is as the
ordinary $ \arctan
\left( \frac{y}{x} \right)$, except that it takes into account in which quadrant
the point $\left( x, y \right)$ is. The resulting director field,
\begin{equation}
\mathbf{n} = \cos \Phi~\mathbf{\hat{u}} + \sin 
  \Phi ~\mathbf{\hat{v}}.
\end{equation}
with
\begin{equation}
\Phi = \alpha+\sum_{i=1,2}\arctan \left( v, u-u_i \right)
\end{equation}
is displayed in Fig. \ref{fig:calc}b for $\alpha=0$. 
We rewrite this by substituting the following identities,
\begin{align}
\mathbf{\hat{u}}=\cos \omega ~\boldsymbol{\hat{\rho}} - \sin \omega ~\boldsymbol{\hat{\omega}}, \\
\mathbf{\hat{v}}=\sin \omega ~\boldsymbol{\hat{\rho}} + \cos \omega ~\boldsymbol{\hat{\omega}},  
\end{align}
where $\omega$ and $\rho$ are the azimuth angle and the radial distance in the $uv$-plane, respectively. The result reads
\begin{equation}
\mathbf{n} = \cos \left( \Phi - \omega \right)  \boldsymbol{\hat{\rho}} 
 + \sin \left(\Phi - \omega \right) \boldsymbol{\hat{\omega}}.
\end{equation}
This field is projected onto a sphere by means of an inverse stereographic projection\cite{1992JPhy2...2..371L}, illustrated in Fig. \ref{fig:stereo}.
Each point on the sphere is represented by a point in the plane by the following relation:
\begin{align}
\label{eq:stereo}
u+iv=2 \mathcal{R} \tan \left( \frac{\theta}{2} \right) e^{i\phi}.
\end{align}
Then, noting that the planar polar unit vectors, ($\boldsymbol{\hat{\rho}}$,
$\boldsymbol{\hat{\omega}}$) are mapped to the spherical ones, ($\boldsymbol{\hat{\theta}}$,
$\boldsymbol{\hat{\phi}}$), we find 
\begin{equation}
\label{eq:ansatz_homogeneous}
\mathbf{n} = \cos \left( \Phi - \phi \right)  \boldsymbol{\hat{\theta}} 
 + \sin \left(\Phi - \phi \right) \boldsymbol{\hat{\phi}}.
\end{equation}
with $\Phi$ given by
\begin{equation}
\Phi = \alpha + \sum_{i=1,2} \arctan \left( \mathcal{Y}_i,\mathcal{X}_i\right) 
\end{equation}
with
\begin{equation}
\mathcal{X}_i = \tan \left( \frac{\theta}{2} \right) \cos \phi - \text{sgn}\left(u_i\right) \tan{\left(\frac{\theta_i}{2}\right)} 
\end{equation}
and
\begin{equation}
\label{eq:Y}
\mathcal{Y}_i = \tan \left( \frac{\theta}{2} \right) \sin \phi
\end{equation}

as the director field on the sphere, depicted in Fig. \ref{fig:splitted sphere}. In ref. \cite{1992JPhy2...2..371L} it was shown that this field on the sphere minimises the free energy in the one-constant approximation. 
It posseses two charge one defects at zenith angles
\begin{equation}
\theta_i = 2\arctan \left( \frac{\left| u_i \right|}{2\mathcal{R}} \right).
\end{equation} 
At the same time this expression is an Ansatz for a homogeneous shell
with two straight disclination lines spanning the shell,
provided we build it out of concentric spheres of radius $\mathcal{R}$.
The director lies along the spheres, including the special case
that these spheres are the surface of the inner or outer
droplet. Therefore, the tangential boundary conditions are satisfied. 

\subsection{An electrostatic analogy and the Ansatz for the inhomogeneous shell}
\label{subsec:electrostatic_analogy}
The concentric spheres that fill up the homogeneous shell are
displaced if the shell is inhomogeneous.
 Moreover, the disclination lines are no longer straight.
 To construct an Ansatz for the director in inhomogeneous shells we
 need to find equations for the displaced spheres and the defect
 lines. For this, we exploit an electrostatic analogy, namely,
 calculating the equipotential (solid in Fig. \ref{fig:electrostatic
   problem}) and electric (dashed green) field lines of an infinitely
 long charged line running parallel to a conducting plane (blue) at a distance $d$. 
\begin{figure}[h]
\centering
\includegraphics[width=0.7\columnwidth]{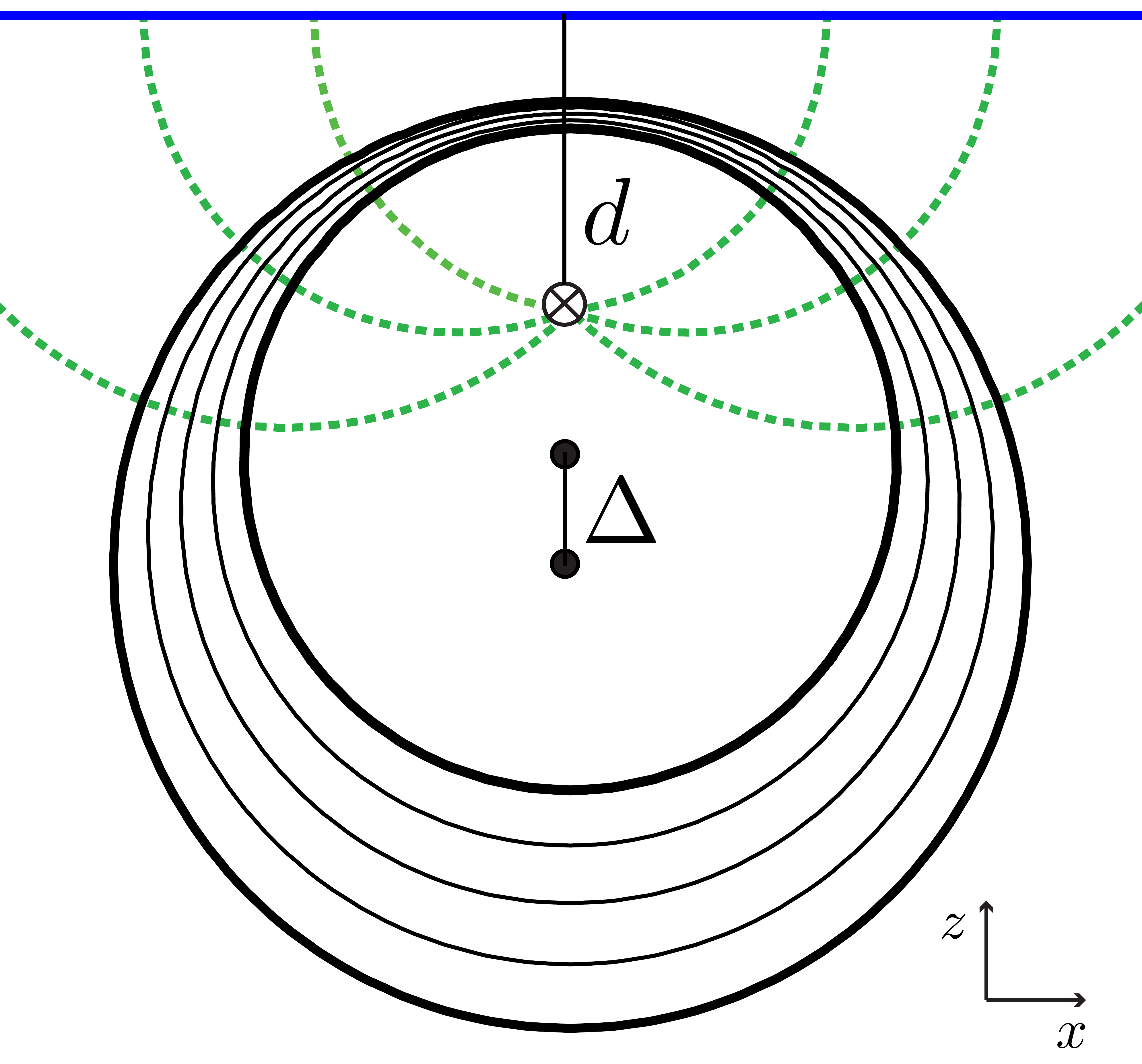}
\caption{Equipotential (solid) and field (dashed green) lines of an infinitely long charged line,
indicated with a cross, running along the y-direction, parallel to a conducting plane (blue) at a
distance $d$. The two equipotential circles, drawn in bold, correspond to two non-concentric
droplets whose centers, indicated by dots, are displaced by $\Delta$.}
\label{fig:electrostatic problem}
\end{figure}
By the method of images, solving this electrostatic problem is equivalent to
solving for the equipotential and electric field lines of two
oppositely charged parallel running cylinders, or, equivalently, a 2D point
charge and its mirror charge. These can be extracted from the complex potential \cite{Smirnov,Hydrodynamics} 
\begin{equation}
\label{eq:map}
\psi \left( w\right) = \log \left( \frac{w+id}{w-id}\right).
\end{equation}
where $w=x+iz$ is a complex number. Note that $\psi$ is a conformal transformation, just like the
inverse stereographic projection is, mapping a region
bounded by two non-concentric circles into a vertical strip (see
appendix \ref{app:mapping}). Thus, the level curves of the real and
imaginairy parts of $\psi\left( w\right)$ are the equipotential and
electric field lines, respectively. These two families of orthogonal
lines, together forming what is known as a isothermic net, read
mathematically 
\begin{align}
\label{eq:real_part}
&\Re{\left[\psi \left( w\right)\right]} = \log \left| \frac{w+id}{w-id}\right| = \text{constant}, \\
\label{eq:im_part}
&\Im{\left[\psi \left( w\right)\right]} = \arg \left( \frac{w+id}{w-id}\right) = \text{constant}.
\end{align}
We see that eq. \ref{eq:real_part} describes circles of Apollonius (see appendix
\ref{app:mapping}) with inverse points $\pm id$. We can rewrite them as
\begin{equation}
\label{eq:circle_A}
x^2+\left(z-\eta\right)^2=\mathcal{R}^2.
\end{equation}
with radius, $\mathcal{R}$, and displacement with respect to the
origin, $\eta$: 
which are related by
\begin{equation}
\label{eq:r_and_eta}
\mathcal{R}^2=\eta^2-d^2.
\end{equation}
Rotating the circular equipotential lines around the $z$-axis creates
the non-intersecting spheres. By choosing two of these spheres (drawn
in bold in Fig. \ref{fig:electrostatic problem}) as the surface of
our inner and outer droplets in addition to a choice of $d$, we can fix the geometry. The relative
displacement of the inner droplet with respect to its concentric
postion, $\Delta$, is given by 
\begin{equation}
\Delta= \eta_{a} - \eta_{R},
\end{equation}
where $\eta_{a}$ and $\eta_{R}$ are the vertical displacements from the origin of the inner and
outer spheres. We take $\eta<0$, such that $\Delta>0$. This implies
that the thinnest part of the shell is at the top, like in Figs.
\ref{fig:electrostatic problem} and \ref{fig:droplet}. 
The other spheres fill up the shell. Since the spheres are the surfaces
of revolution of the circles around the $z$-axis, we obtain the equation for the
spheres simply by addition of $y^2$ to the left hand side eq. (\ref{eq:circle_A}):
\begin{equation}
\label{eq:sphere}
x^2+y^2+\left(z-\eta\right)^2=\mathcal{R}^2.
\end{equation}
Two independently chosen electric field lines
will serve as disclination lines. These 
lines run perpendicular to the equipotential lines, and thus
perpendicular to the surface of the inner and outer droplet, as is
demanded by the tangential boundary conditions. Similar to the
calculation of the equipotential lines, one can obtain the equations
for the electric field lines from eq. \eqref{eq:im_part} (appendix \ref{app:mapping}). We find that the electric field lines are also circles:
\begin{equation}
\label{eq:electric_field_lines}
\left(x-\epsilon\right)^2+z^2=S^2,
\end{equation} 
with radii, $S$, and displacements, $\epsilon$, now in the
$x$-direction, which are related as follows:
\begin{equation}
S^2=\epsilon^2+d^2.
\end{equation}
Since only the circular arc that is inside the shell matters, we care
about the points of intersection of the two defect lines with the spheres that fill up the
shell. We would like to find the zenith angle on each
sphere, $\beta_i$ , that these points of intersection make. 
We assign a different character than $\theta_i$, because $\beta_i$
does not have a constant value as it depends on the displacement (or radius) of
the sphere. Let us therefore
refine our definition of $\theta_i$ as the zenith angle of the
defect on the outer-most sphere.
Now, the following geometrical relations hold:
\begin{align}
x&=\pm \mathcal{R} \sin \beta_i \\
z&=\eta+\mathcal{R} \cos \beta_i
\end{align}
Then, by substituting $x$ and $z$ in the eq. (\ref{eq:electric_field_lines})
and eliminating $d$ in favor of $\mathcal{R}$ by
applying eq. \eqref{eq:r_and_eta} we find an expression for
$\epsilon_i$ as a function of $\beta_i$, $\eta$ and $\mathcal{R}$
(assuming $\sin \beta_i \neq 0$):
\begin{equation}
\label{eq:epsilon}
\epsilon_i=\pm\frac{\mathcal{R}+\eta\cos\beta_i}{\sin\beta_i} = \pm\frac{\mathcal{R}+\eta\cos\theta_i}{\sin\theta_i}
\end{equation} 
where the last equality follows from the constantness of
$\epsilon_i$, as we are moving along the same circle.
We find the solution for $\beta_i$ 
\begin{equation}
\label{eq:beta}
\beta_i=2\arctan\left( \frac{\epsilon_i+\sqrt{\epsilon_i^2+d^2}}{\mathcal{R}-\eta} \right)
\end{equation} 
Not surprisingly, $\beta_i$ is increasing as the radius of the sphere
is decreasing. If $\theta_i=0$ or $\theta_i=\pi$, the disclination lines are straight and $\beta_i=0$ or $\beta_i=\pi$, respectively. Next, we find $\eta$ as a function of the spatial
coordinates $x$, $y$ and $z$, since it is the only variable, besides
the parametric dependence on $d$ and $\theta_i$, on which $\beta_i$ is depending. Put 
differently, given some point in space, on which sphere is it? To
answer this question we resort to eq. \eqref{eq:sphere}, yielding the
following result:
\begin{equation}
\label{eq:eta}
\eta \left( \mathbf{x} \right) = \frac{x^2+y^2+z^2+d^2}{2z}.
\end{equation}
We have now acquired all the necessary information to construct the
Ansatz for the director field in an inhomogeneous shell. We take the
Ansatz for the director field in a homogeneous shell, eq.
\eqref{eq:ansatz_homogeneous}, and make the following replacements 
\begin{align}
\theta_i \rightarrow \beta_i, \\
z \rightarrow z - \eta
\end{align}
The first substitution concerns the defect lines. The second accounts
for the displacement of the spheres and implies the substitution
\begin{align}\
\label{eq:substitution}
\theta \rightarrow \beta = \arccos \left( \frac{\left(z -\eta \right)}{x^2+y^2+\left(z -\eta \right)^2} \right), 
\end{align}
with $\beta$ being the zenith angle on the displaced sphere.
Finally, together with eqs. \eqref{eq:epsilon}-\eqref{eq:substitution} we obtain the Ansatz for the director in inhomogeneous shells
with two charge one disclination lines:  
\begin{equation}
\label{eq:ansatz_inhomogeneous}
\mathbf{n} = \cos \left( \Phi - \phi \right)  \boldsymbol{\hat{\beta}} 
 + \sin \left(\Phi - \phi \right) \boldsymbol{\hat{\phi}}.
\end{equation}
where $\Phi$ is now given by
\begin{equation}
\Phi = \alpha + \sum_{i=1,2} \arctan \left( \mathcal{Y}_i,\mathcal{X}_i\right) 
\end{equation}
with
\begin{equation}
\mathcal{X}_i = \tan \left( \frac{\beta}{2} \right) \cos \phi - \text{sgn}\left(u_i\right) \tan{\left(\frac{\beta_i}{2}\right)} 
\end{equation}
and
\begin{equation}
\label{eq:Y}
\mathcal{Y}_i = \tan \left( \frac{\beta}{2} \right) \sin \phi
\end{equation}
The disclination lines can be put anywhere except for the south pole. In the case of a
bipolar defect arrangement, i.e. $\theta_i=0$ and $\theta_i=\pi$, we draw on each sphere the director given by eq. (\ref{eq:ansatz_homogeneous_bipolar}), with the substitution in eq. \eqref{eq:substitution}
and find an ansatz for the bipolar inhomogeneous shell that reads
\begin{equation}
\mathbf{n} \left(\mathbf{x}\right) = \boldsymbol{\hat{\beta}} =\cos \beta \cos \phi
\mathbf{\hat{x}} + \cos \beta \sin \phi \mathbf{\hat{y}} - \sin \beta \mathbf{\hat{z}}.
\end{equation}   

\begin{figure}[h]
\centering
\includegraphics[width=0.95\columnwidth]{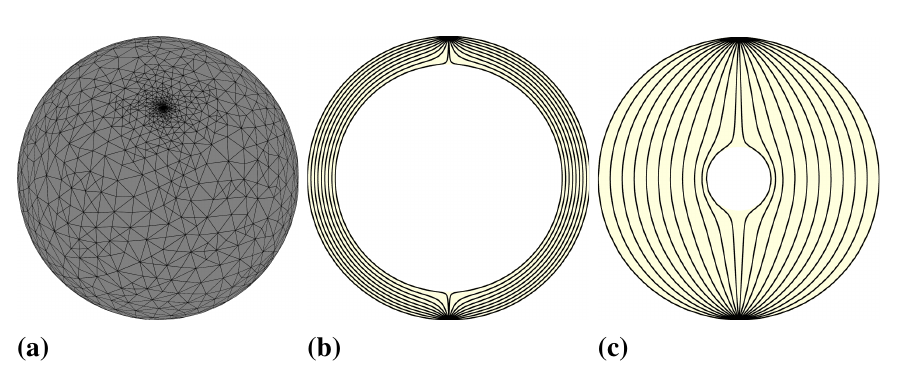}
\caption{(a) A typical mesh used in the numerical minimisation is refined around the locations of the defects. (b,c) Director field of the shell for (b) $\frac{h}{R}=0.2$ and (c) $\frac{h}{R}=0.77$. The disclination lines escape in the third dimension leaving two pairs of boojums. Each pair is encircled in red in (b) The director field in (c) resembles a slightly distorted director field of a single nematic droplet.}
\label{fig:director2}
\end{figure}

The Ansatz is then subjected to a numerical minimisation, employing the finite element method \cite{1999EPJB...10..515S} suitable for non-trivial geometries, to ensure the escape of the disclination lines leaving a point defect at the inner and outer surface for each line (see Figs. \ref{fig:director} and \ref{fig:director2}). We refine the mesh at these defects to obtain good accuracy on the rapidly changing director (fig. \ref{fig:mesh}).

\section{Energetics of homogeneous shells}
\label{sec:homogeneous}
\begin{figure}[h]
\begin{center}
\includegraphics[width=\columnwidth]{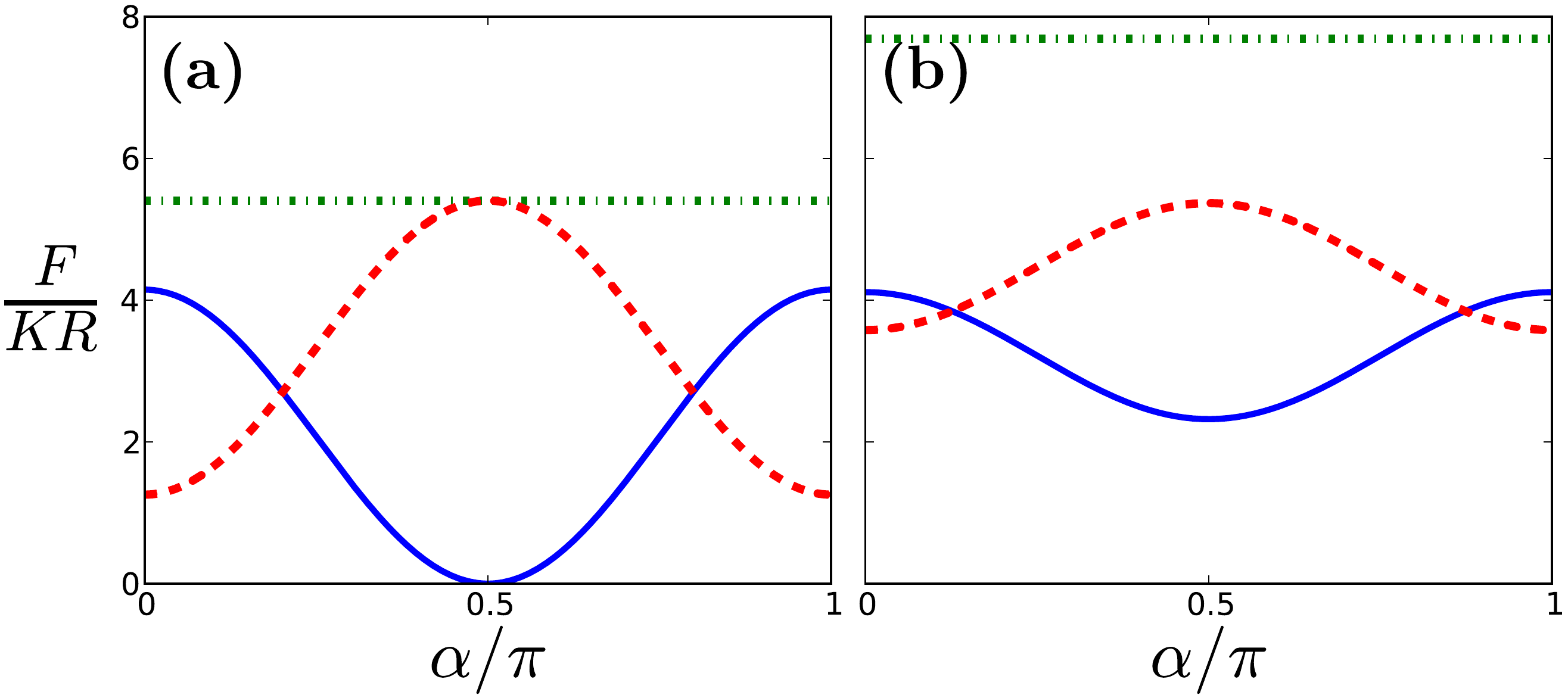}
\caption{The splay (solid blue), bend (dashed red), and their sum (dashed-dotted green) as a function of $\alpha$ when (a) $\theta_{12}=\pi$ and (b) $\theta_{12}=0.1\pi$, in both cases $\frac{b}{R}=0.025$.}
\label{fig:anisotropy}
\end{center}
\end{figure}

\begin{figure}[h]
\begin{center}
\includegraphics[width=0.7\columnwidth]{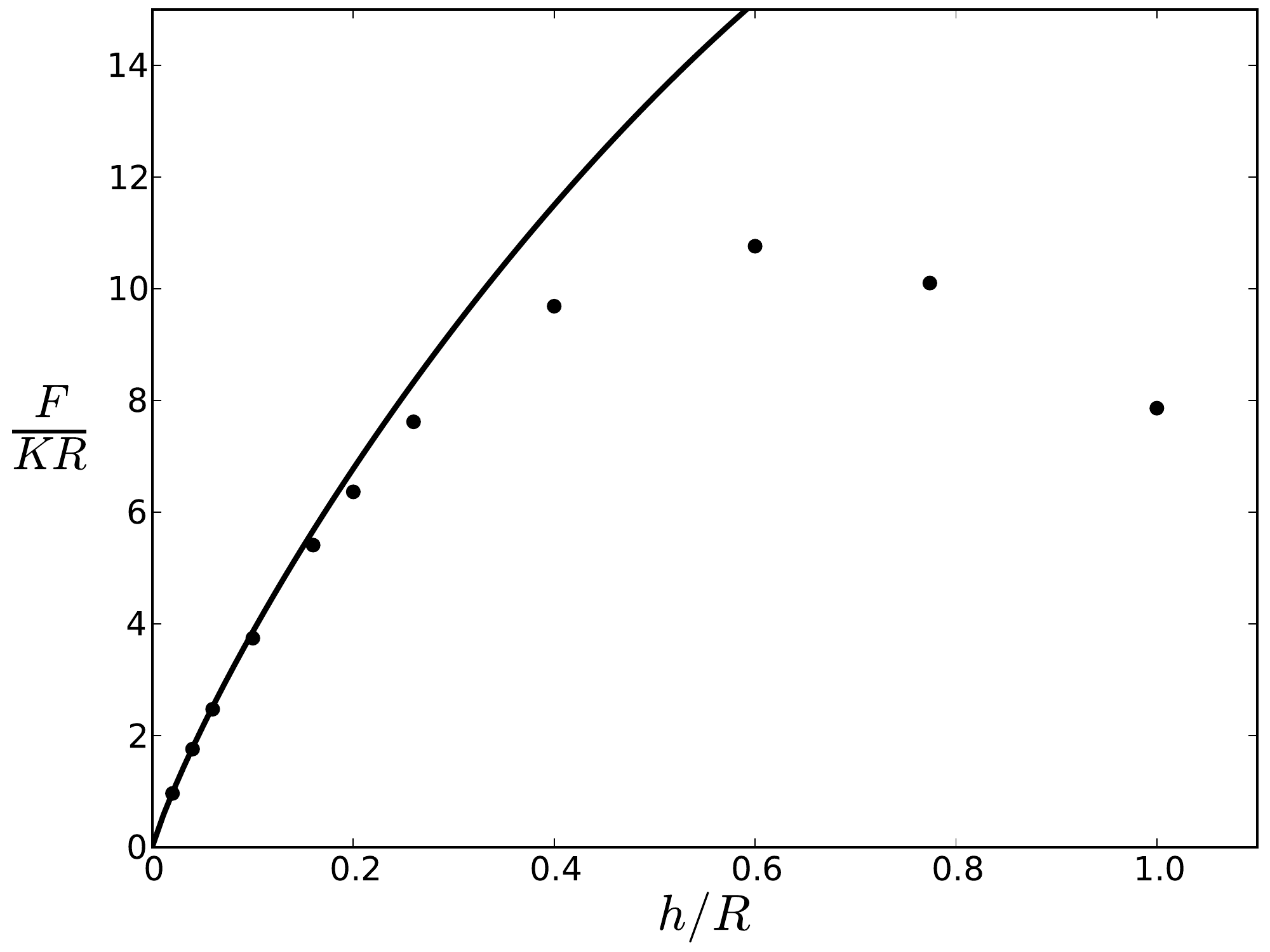}
\caption{Free energy of the bipolar shell as a function of thickness. The line is given by the analytical estimate in eq. \eqref{eq:energy_vs_thickness}.}
\label{fig:energy_vs_thickness}
\end{center}
\end{figure}

First, we calculate the free energy for the Ansatz in
\eqref{eq:ansatz_homogeneous_bipolar}, in which the defects are located on opposite
poles on the sphere, by integrating the free energy density over the spherical shell except for a cut-off region determined by $0 < \theta < \frac{b}{r}$ and $\pi - \frac{b}{r} < \theta < \pi $ with $r$ the radial coordinate. The result reads
\begin{align}
F =& 2 \pi \left( K_1 \cos^2 \alpha + K_3 \sin^2 \alpha \right) \\ &\left( R \log{\frac{2R}{b}} - \left( R - h \right)
  \log{\frac{2 \left( R - h \right)}{b}} -2h \right) \\&+ 2 \pi \left(K_3 - 2 K_{24} \right) h  \\
\xrightarrow{K_i=K}&   2 \pi K \left( R \log{\frac{2R}{b}} - \left( R - h \right)
 \log{\frac{2 \left( R - h \right)}{b}} -2h \right)
\end{align}
Note that splay and bend deformations can be transformed into each
other by tuning $\alpha$, but the total energy is
unchanged if $K_1=K_3$, as is shown graphically in Fig. \ref{fig:anisotropy}a.
Furthermore, note that the saddle-splay term is
proportional to the thickness, in contrast to the twist and bend
energy which both contain a logarithmic divergence. We therefore
expect that, as a first approach, it is not so important in determining the defect locations. Since
these defects repel each other, this director field is expected to be the
ground state. Upon placing the defects at a different separation from
each other, eq. \eqref{eq:ansatz_homogeneous}, leads to an increase in the elastic energy\cite{1992JPhy2...2..371L,2002NanoL...2.1125N,2006PhRvE..74b1711V}.
Moreover, the splay and bend cannot be efficiently transferred into one
another by a global rotation (changing $\alpha$), e.g. splay no longer vanishes
for $\alpha =0$ whereas it did for eq. \eqref{eq:ansatz_homogeneous_bipolar}. This is presented graphically in Fig. \ref{fig:anisotropy}b. Note that the director field minimising the free energy for $K_1 \neq K_3$ is not equal to the Ansatz \cite{2008JChPh.128j4707B,2008PhRvL.101c7802S,PhysRevE.62.5081,2011PhRvL.106x7802L,2011PhRvL.106x7801L}. Besides the elastic anisotropy the escape of the defect lines in the third dimension modifies the energetics. As a result, there are two pairs
of boojums residing on the interfaces. We can effectively take the escape into
account in our calculations of the energy by replacing the cut-off $b$
by the thickness $h$ and adding $4.2 \pi K h$ for each pair of boojums\cite{2006PhRvE..74b1711V,2002PhRvE..66c0701C}. We obtain in the one-constant approximation
\begin{equation}
\label{eq:energy_vs_thickness}
F = 2 \pi K \left( R \log{\frac{2R}{h}} - \left( R - h \right)
  \log{\frac{2 \left( R - h \right)}{h}} + 2.2h \right) 
\end{equation}
In Fig. \ref{fig:energy_vs_thickness} we compare this analytical estimate with numerical results from our
procedure outlined in the previous section. We find a good agreement,
in particular for small $\frac{h}{R}$, as expected.
In this regime, the
free energy rises as the volume of the shell increases. For large $\frac{h}{R}$, our result deviates from eq. \eqref{eq:energy_vs_thickness}. Remarkably, as
the thickness is increased, the free energy decreases after some critical value,
$\frac{h}{R} \approx 0.6$. The size of the inner droplet, which is $2a$, is no longer larger that the
scale over which the escape happens, which is roughly $h$.  As a result, when $h$ becomes comparable to $R$, the
inner droplet no longer forms
an obstruction that makes the shell locally look like a slab in which
the lines can escape. Rather, the point of view that a slight director distortion is induced by a single nematic
droplet (resulting in an energy cost) is more appropriate in this regime.
This cross-over is illustrated in Figs. \ref{fig:director} and \ref{fig:director2}.  
So, for thin shells, it is energetically favorable to decrease the distance
between the boojums that form a pair. However, for thick shells, the opposite can be concluded: it is energetically
favorable to increase the intrapair distance between boojums. 
Apart from the \textit{intrapair} interaction described above, 
there is an \textit{interpair} interaction. This is repulsive in nature, thus irrespective of thickness, 
for homogeneous shells we always find the bipolar arrangement 
as the free energy minimum. 

\section{Energetics of inhomogeneous shells}
\label{sec:inhomogeneous}
\subsection{Buoyancy versus elastic forces}
\begin{figure}[h]
\begin{center}
\includegraphics[width=\columnwidth]{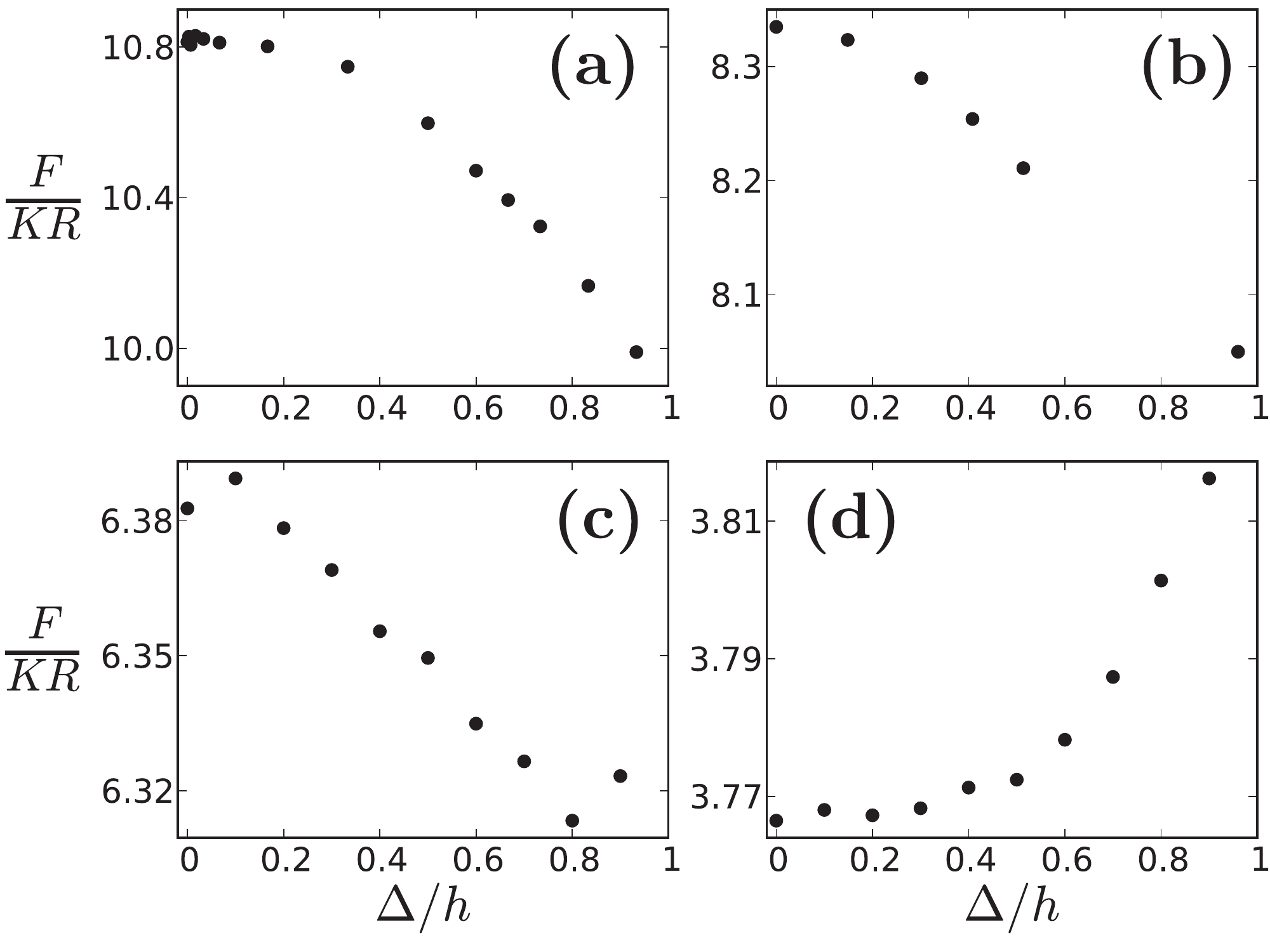}
\end{center}
\caption{The elastic free energy of the liquid crystal as a function
  of the relative displacement of the inner droplet. In (a), (b)
  and(c),$\frac{h}{R}=0.6,0.3,0.2$, respectively,  the energy is
  minimised when the droplet is on the periphery of the larger
  droplet, resulting in an elastic force of the order of $K$
  pushing the inner droplet outwards. (d)   
For a thinner shell with $\frac{h}{R}=0.1$, there is a restoring force on
the inner droplet, driving it back to the center of the outer one. }
\label{fig:energy_vs_displacement}
\end{figure}

Before we study the effect of the thickness inhomogeneity on the
mechanics of the nematic liquid crystal, we first investigate its
origin. In our experiments we observe that the inner water droplet is
displaced along the vertical direction. This implies that
gravity plays its part, but it does not necessarily mean that it is
the density mismatch between the nematic and water that drives the
motion of the inner droplet. Another possibility would be that the elastic
forces push the droplet out of the center, while gravity only breaks
the symmetry. To identify the origin of the thickness ihhomogeneity, we will compare the magnitude of
the elastic forces with Archimedes force. Therefore we map out the
elastic energy as a function of the displacement $\frac{\Delta}{h}$ 
for several values of $\frac{h}{R}$, as shown in Fig. \ref{fig:energy_vs_displacement}.  
Our first observation is that the stability of this perturbation is a
nontrivial function of the thickness. For $\frac{h}{R}=0.2,0.3,0.6$ we
observe that the energy decreases as a function of $\Delta$.  This is in
agreement with a calculation in  ref. \cite{2007PhRvL..99o7801F}
done for $\frac{h}{R}=0.77$. However, for a relative thin shell of
$\frac{h}{R}=0.1$ there is an elastic minimum for $\Delta = 0$.
Second, the magnitude of the elastic force is less than or of the order of
$f_{e} \sim K \approx 10^{-11} N$. This is much smaller than the
net force from buoyancy and the weight of the droplet
$f_b=\left( \rho_{nem} - \rho_{w}\right) g V$ with the volume of the
water droplet $V=\frac{4}{3}\pi a^3$. For $a\approx 50~\mu m$ and 
a difference in density between 5CB and water of roughly $3~10~kg~m^{-3}$ at room temperature\cite{doi:10.1021/jp711211w}, we find
$f_b\approx 2~10^{-10} N$. Therefore, we conclude that  buoyancy is indeed responsible for displacing the inner water droplet from the center in our experiments. If one would try to match the density of the
nematic to the water density, as was done in the experiments in ref. \cite{2007PhRvL..99o7801F}, where the density difference was brought down to $2~
kg ~m^{-3}$, $f_b$ and $f_{e}$ will be of the same order, but only when the inner droplet is at the periphery. Also in the regime of small $a$ these
forces will become comparable. 

\subsection{Confined and deconfined defect configurations}
\begin{figure}[h]
\begin{center}
\includegraphics[width=\columnwidth]{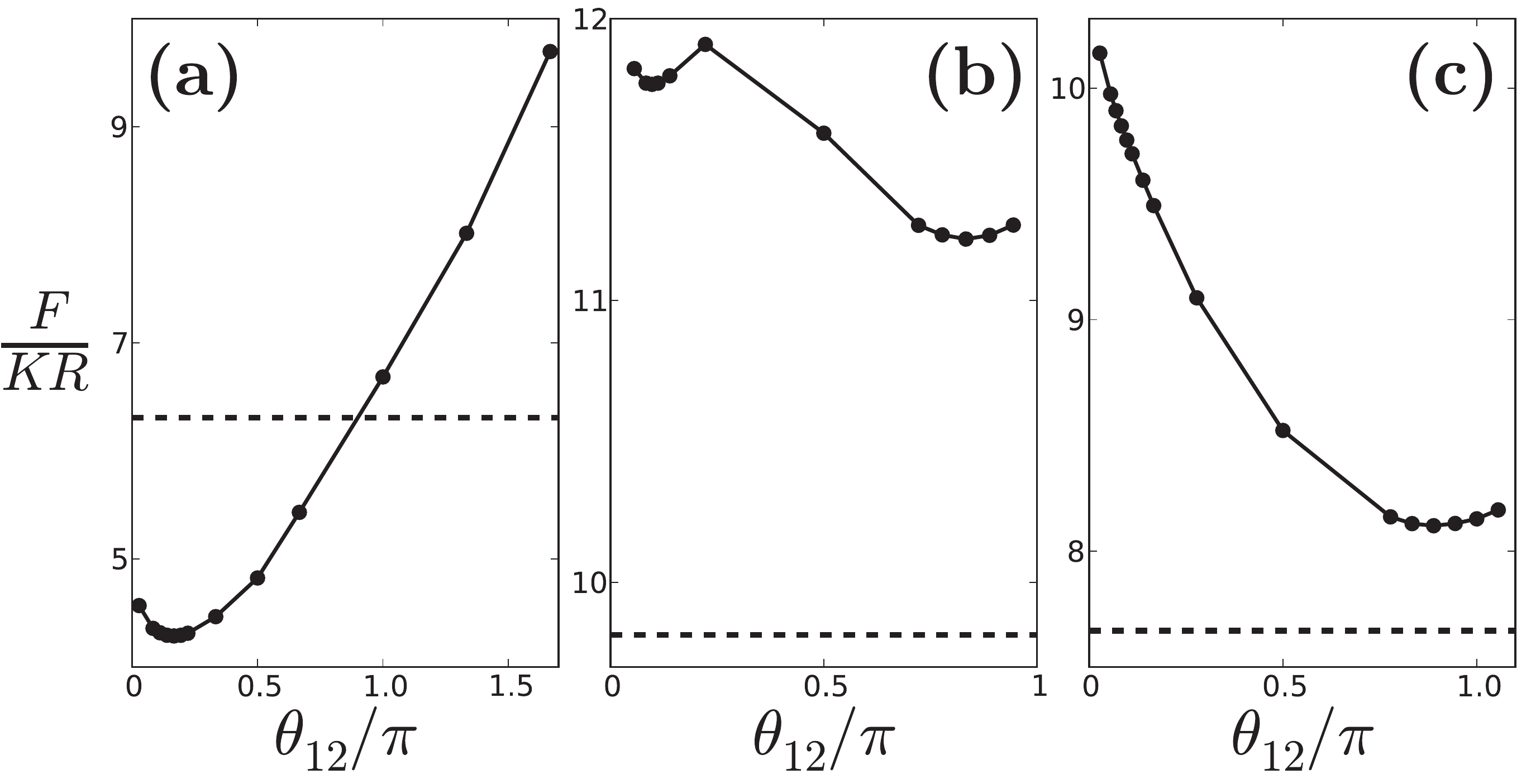}
\caption{The free energy for a shell as a function of the angular separation between the two defects on the outer surface in the confined configuration, when (a) $\frac{h}{R}=0.2$ and $\frac{\Delta}{h}=0.831$ (b) $\frac{h}{R}=0.7$ and $\frac{\Delta}{R}=0.939$, (c) $\frac{h}{R}=0.8$ and $\frac{\Delta}{h}=0.946$.  The dashed line indicates the energy of the deconfined configuration. These graphs suggest a confined global minimum at $\frac{\theta}{\pi}\approx0.17$ in (a), a local confined minimum at $\frac{\theta}{\pi}\approx0.1$ and a global deconfined minimum in (b) and a global deconfined minimum in (c). }
\label{fig:energy_vs_angle}
\end{center}
\end{figure}

In the remainder of this article we will compare two distinct defect configurations. In one 
configuration the defects are at maximum angular separation from 
each other at opposite sites on the sphere. This we will refer to as the deconfined
state. In the other case the defects are trapped or confined to the
thinnest top part of the shell. The defects are located symmetrically at an angle $\theta_i$ from the vertical axis so that their angular separation is simply $\theta_{12}=2\theta_i$. 
The energy can be estimated to grow with the thickness of the shell where the defects are located. This is roughly the minimal thickness at the top of the shell, for which there is a simple geometrical relation $h_{min}=h - \Delta$. From this one immediately sees that $h$ and $\Delta$ take opposite roles. 
We thus expect the confined state
to be energetically favorable over the deconfined state when the shell
is sufficiently thin and inhomogeneous, \textit{i.e.} low $h$ and high $\Delta$. This heuristic argument has led us
to a systematic study of the energy landscape as a function of defect location. We classify three cases: I)
the deconfined state is the only energy minimum , see Fig. \ref{fig:energy_vs_angle}a ; II) both the confined and deconfined state are minima, one of them is local and the other is global, see Fig. \ref{fig:energy_vs_angle}b ; III) the confined state
is the only energy minimum, see Fig. \ref{fig:energy_vs_angle}c. 

\subsection{Phase diagram}
\begin{figure}[h]
\begin{center}
\includegraphics[width=\columnwidth]{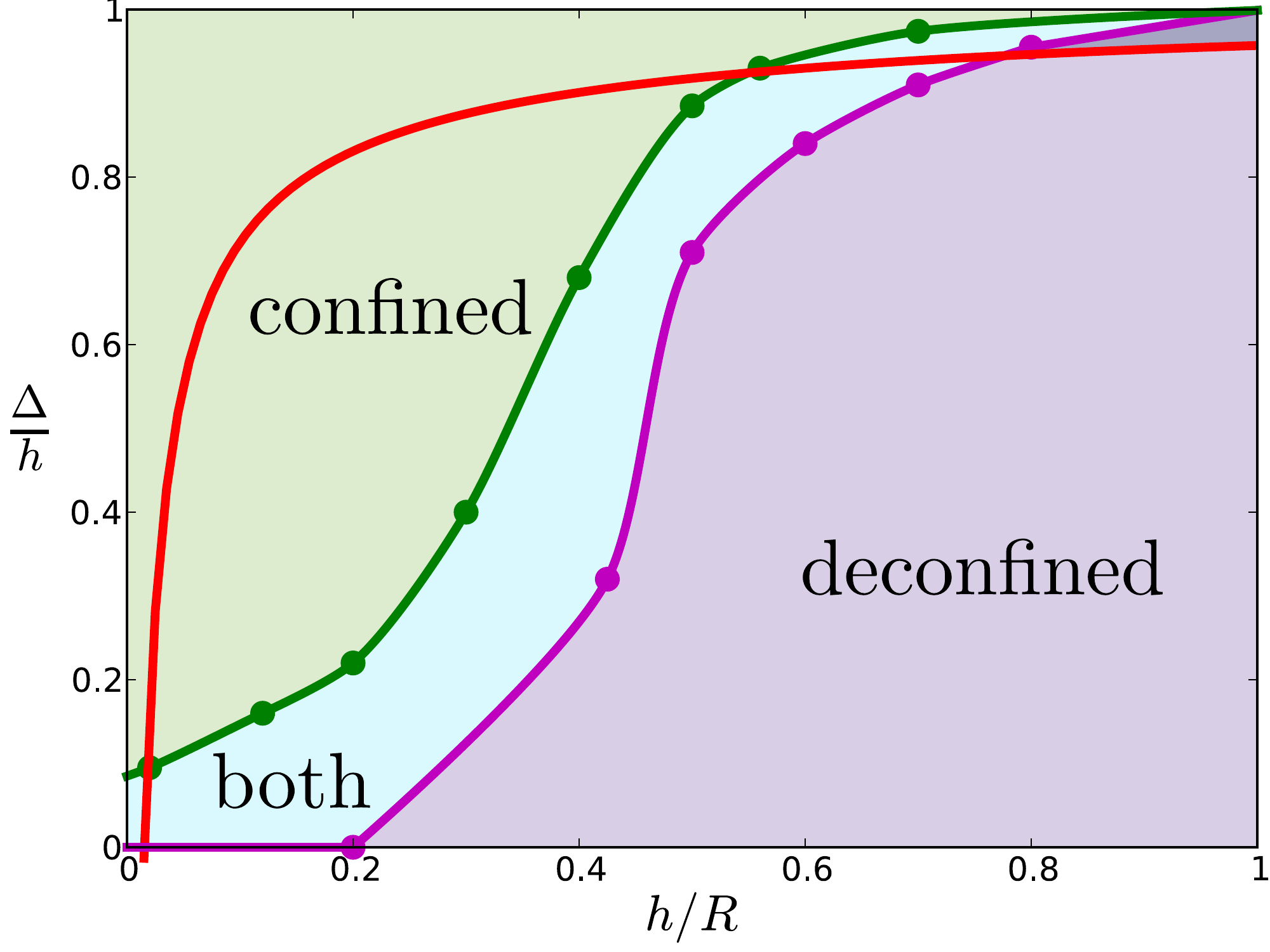}
\caption{Phase diagram of the confined (green), deconfined (purple) and coexistentance phase (blue) as a function of thickness of the shell, $\frac{h}{R}$ and thickness inhomogeneity $\frac{\Delta}{h}$. The confinement (green), $\frac{\Delta_c}{h}$, and deconfinement (purple), $\frac{\Delta_d}{h}$, transition lines separate these phases. The red line represents the assumed experimental trajectory of constant $h_{min}$.}
\label{fig:phase_diagram}
\end{center}
\end{figure}

We construct a phase diagram as a function of thickness and thickness
inhomogeneity. We find that for a given thickness there is a deconfined minimum below a critical value of the relative displacement $\frac{\Delta_c}{h}$, marked green in Fig. \ref{fig:phase_diagram}, which is monotonously increasing with the thickness. 
The confined state is found to minimise the energy above another critical value, $\frac{\Delta_d}{h}$, marked in purple in Fig. \ref{fig:phase_diagram}, which is also larger for thicker shells.  Therefore, as anticipated in the previous section, we find that the confined defect state minimises the elastic energy for thin and inhomogeneous shells, whereas the deconfined defect state minimises the energy for rather homogeneous and thick shells. Since these two critical values for $\frac{\Delta}{h}$ are different there exist two minima for $\frac{\Delta_d}{h} < \frac{\Delta}{h} < \frac{\Delta_c}{h}$. We can thus divide the phase diagram in three regions: a deconfined minimum-only, confined minimum-only and coexisting region coloured purple, green and blue in Fig. \ref{fig:phase_diagram}, respectively. These phases are separated by lines marking where, as in a first-order phase transition, a local energy minimum is lost. We remark that the energy differences between the
deconfined and weakly confined states for thin and homogeneous shells
become too small to conclude with certainty that $\frac{\Delta_c}{h}$ goes to a finite value and the deconfinement transition
reaches $\frac{\Delta_d}{h}=0$ at extremely low $h$. 

\subsection{Comparison with experiment}
The nematic double emulsion droplets create inhomogeneous shells,
because buoyancy displaces the inner droplet upward from its concentric
position along the gravitational direction. The short-range steric
repulsion from the polymer polyvinyl alcohol (PVA), prevents the inner droplet from coalescing with the continuous phase. Therefore, we assume that the thinnest part of
the shell, $h_{min}$, is effectively constant. By osmosis the
thickness inhomogeneity can be modified. We find 
\begin{equation}
\label{eq:exppath}
\frac{\Delta}{h} = 1- \frac{u_{0}}{u}  \sqrt[3]{ \frac{1 - \left( 1- u\right)^3}{1 - \left( 1- u_0\right)^3} } 
\end{equation}
where $u \equiv \frac{h}{R}$ and $u_0$ is the value of $u$ when the shell becomes homogeneous, see appendix \ref{app:exppath}. This path through the phase diagram is indicated in red in Fig. \ref{fig:phase_diagram}.
 If we traverse this path in the direction of decreasing thickness we find that the angular separation between the defects, $\theta_{12}$,  changes abruptly from $\pi$ to a value much smaller than that, as does the order parameter in a first-order phase transition. 
\begin{figure}[h]
\begin{center}
\includegraphics[width=\columnwidth]{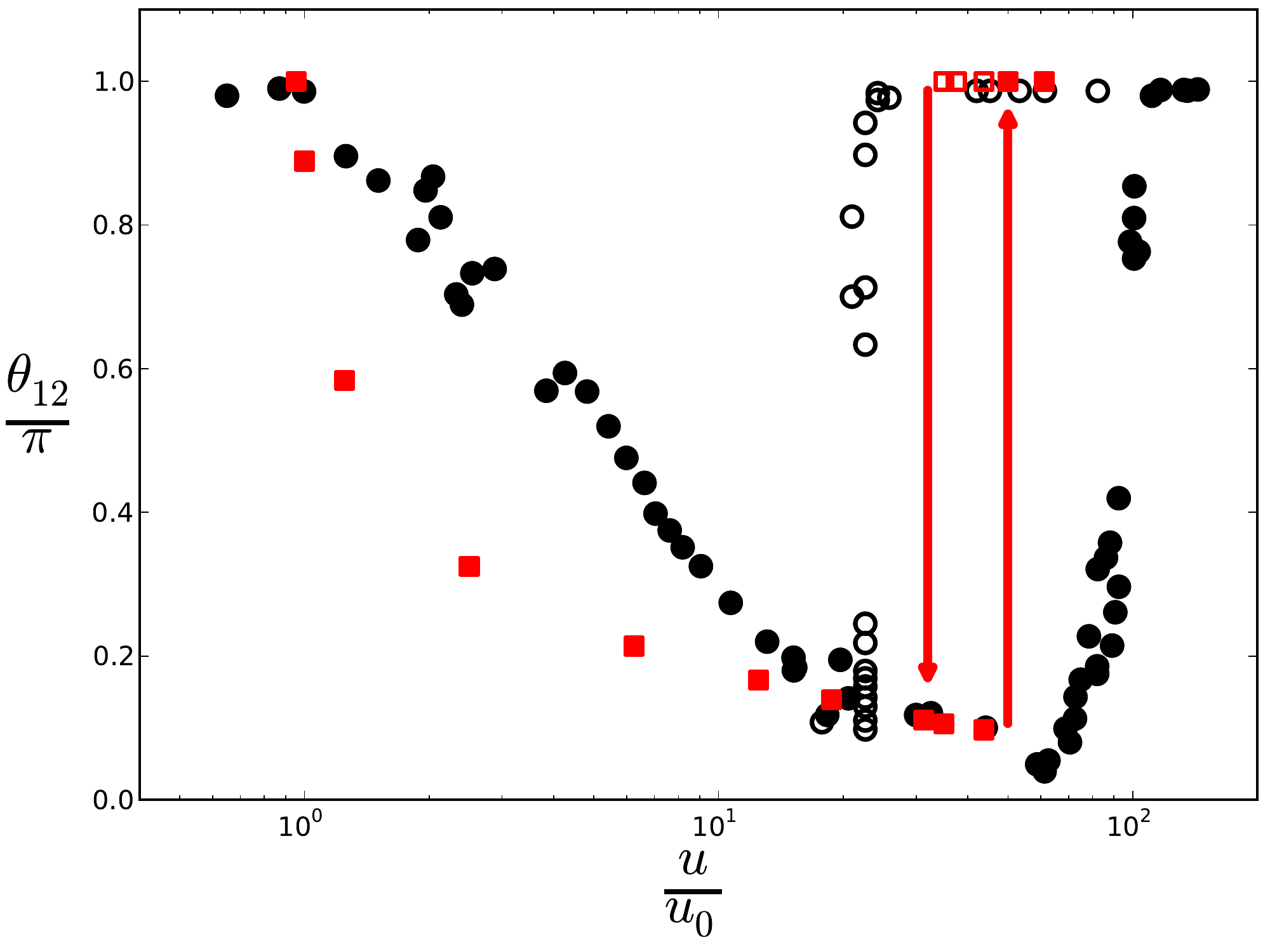}
\caption{Angular separation between the defects as a function of normalised thickness, $\frac{u}{u_0}$, in the experiment (black circles) and in theory (red squares). The open symbols depict the hysteresis.}
\label{fig:comparison}
\end{center}
\end{figure}
In the model this occurs in both theory (red squares in Fig. \ref{fig:comparison}) at $u / u_0 \approx 30$ and in the experiment (black circles in Fig. \ref{fig:comparison}) at $u / u_0 \approx 20$.
The abruptness of the confinement transition is marked by the the short timescale of only tens of seconds, compared to the hours over which the osmosis occurs, in which the pair of defects located at the thicker hemisphere moves toward the top of the shell (see Fig. \ref{confinement_transition_exp}). 
\begin{figure}[h]
\begin{center}
\includegraphics[width=\columnwidth]{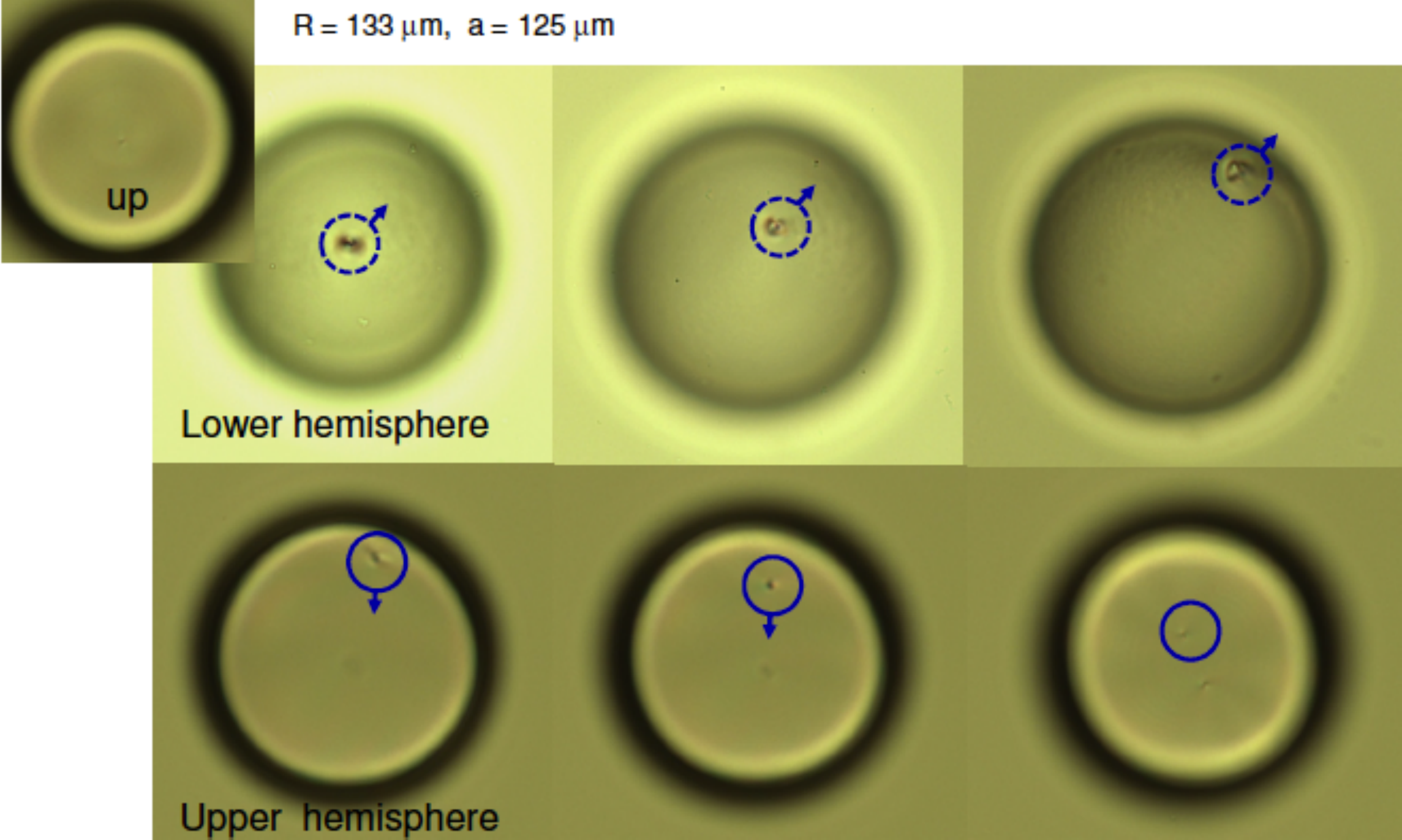}
\caption{Bright field images of a shell (top view) undergoing the confinement transtition. A pair of boojums moves from the lower hemisphere (top row) to the upper hemisphere (bottom row).The time span is tens of seconds. $2R=133~\mu m$ and $2a=125~\mu m$.}
\label{confinement_transition_exp}
\end{center}
\end{figure}
Upon decreasing the thickness and consequently the thickness inhomogeneity even further the defects spread and the angular separation increases again. When the shell is approximately homogeneous (see Fig. 14a), the effect of confinement has weakened so much that the defects are aligned antipodally. The axis joining them can now point in any direction though, as shown by the two shells in Figs. 14b and 14c. In this case, the energy of the thin shell does not depend on the orientation of this axis, in contrast to what happens for thicker shells, whose boojums axis are aligned along the gravitational direction. This also confirms that the defect deconfinement transition in the phase diagram goes to $\Delta = 0$ for low $h$. 
\begin{figure}
{\includegraphics[width=0.95\columnwidth]{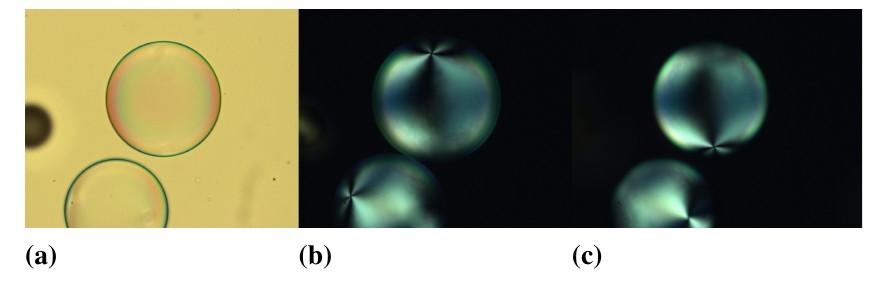}
\label{fig:hello}}
\caption{(a) Bright field image of two thin shells obtained after osmotically shrinking the corresponding thicker shells. The location of the defects are easily seen in cross-polarization by (b) focusing the shells at the top, to see the upper pair of boojums, and (c) at the bottom, to see the lower pair of boojums. The defects are a diameter away, with $\theta_{12} = \pi$. Interestingly, for these thin shells, the direction of the axis joining the two pairs of boojum is not correlated with the gravitational z-axis. The dimensions of the upper shell are $2R =138.8~ \mu m$ and $2a= 137.5 ~\mu m$, corresponding to $u/u_0=1.05$.}
\end{figure}
Upon reversing the path through the phase diagram, \textit{i.e.} traverse the red path in Fig. \ref{fig:phase_diagram} in the direction of increasing thickness, we first find that the defects move toward each other gradually. Upon increasing the thickness even further we find that $\theta_{12}$ increases rapidly to its maximum possible value at $\frac{u}{u_0} \approx 50$ and $\frac{u}{u_0} \approx 80$ in the model and experiment, respectively, as the mutual repulsion between the pairs of defects becomes too large. It is thus favorable to have one pair of boojums at the thickest part of the shell. Note that the thickness at which this deconfinement transition occurs is thus larger than the thickness at which the confinement transition occurs. This hysteresis between the confinement and deconfinement transitions is due to phase coexistence. The green and purple curves in Fig. \ref{fig:phase_diagram} (corresponding to the confinement and deconfinement transition) intersect the red curve (assumed experimental path) at different points in the phase diagram.

\section{Conclusion}
\label{sec:conclusion}
In this study we have crossed from a two-dimensional description of a spherical nematic liquid crystal to a spherical bivalent shell with a finite thickness and possible inhomogeneity. Irrespective of thickness, we always find an antipodal arrangement as the free energy minimum in homogeneous shells of nematic liquid crystal. However, this scenario changes when the shell thickness is sufficiently inhomogeneous. The repulsion between the pairs of boojums competes with the minimisation of the distance between the defects within a pair. As a result, the defects undergo a confinement transition to the thinnest part of the shell. Conversely, the defects confined in the thinnest hemisphere make a deconfinement transition that maximises their separation. The critical displacement of the inner droplet for which these transitions occur are in general not equal, i.e. there is hysteresis present. These transitions are also present in our experiment, where a water droplet encapsulates a nematic liquid crystal droplet to make a spherical nematic shell. We have showed that these shells are inhomogeneous due to the buoyancy that displaces the inner droplet along the gravitational direction. Additional to the confinement and deconfinement transitions, a continuous evolution is observed, when thin shells become less inhomogeneous. Though we found an excellent qualitative agreement between theory and experiment for all these phenomena, an exact quantitative agreement is still lacking, possibly due to a lack of validity of the one-constant approximation. It would be interesting to extend this study by investigating the role of elastic anisotropy on the defect transitions in nematic shells. Since it is more difficult to exchange splay and bend when the defects are confined,  we expect that the region in the phase diagram occupied by the confined state  will be smaller if elastic anisotropy is included. This would imply that the confinement and deconfinement transitions occur at smaller thickness if $K_1 \neq K_3 $.

\section*{Acknowledgement}
V.K. acknowledges funding from Stichting Fundamenteel Onderzoek der Materie (FOM). A. F-N. thanks the National Science Foundation for Career Award DMR-0847304.

\appendix
\section{Conformal mappings and the circles of Apollonius}
\label{app:mapping}
\begin{figure}[htbp]
\begin{center}
\includegraphics[width=\columnwidth]{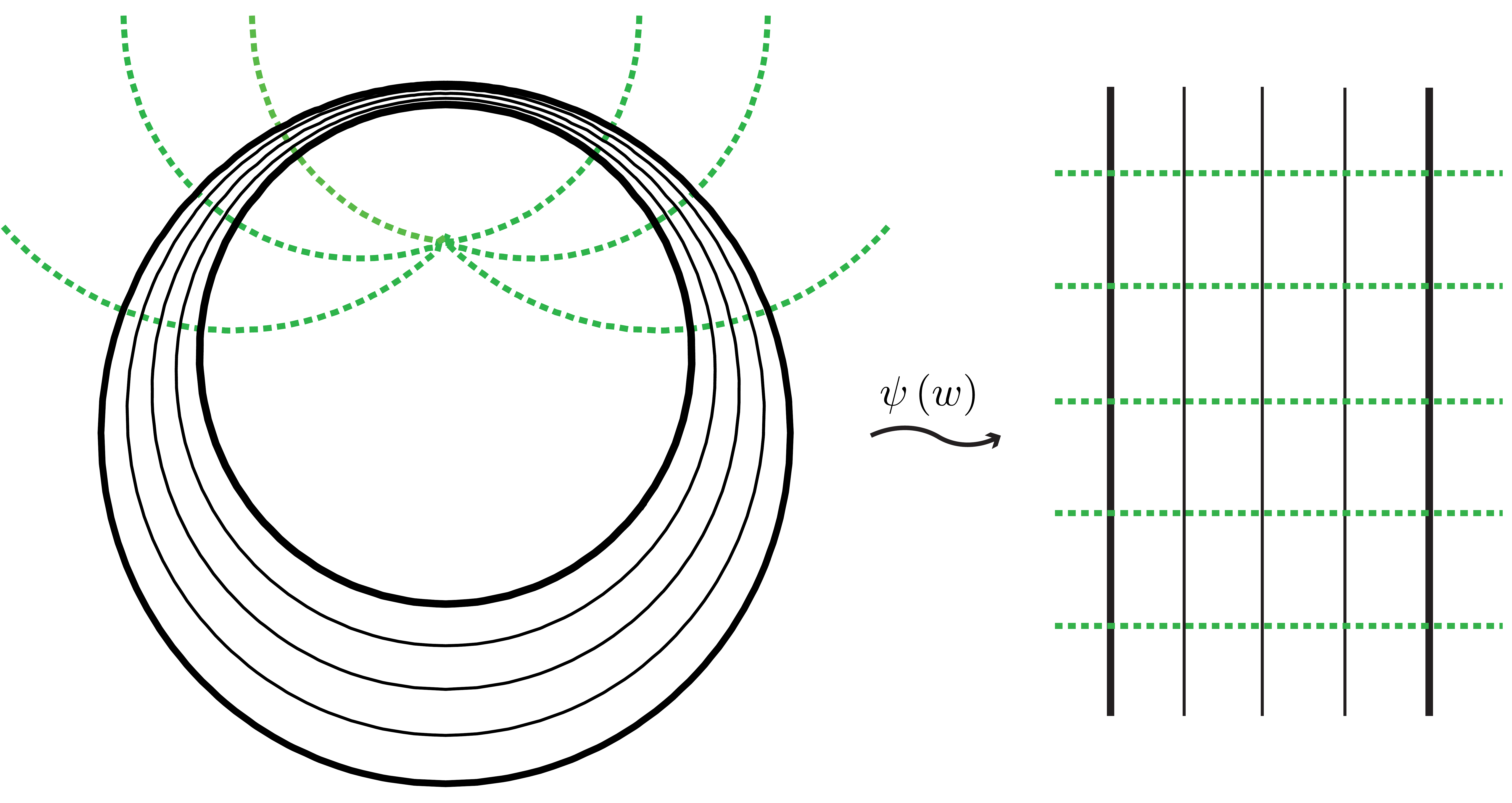}
\caption{Schematic of the mapping of eq. \eqref{eq:map2}.}
\label{fig:map2}
\end{center}
\end{figure}

\begin{figure}[htbp]
\begin{center}
\includegraphics[width=0.5\columnwidth]{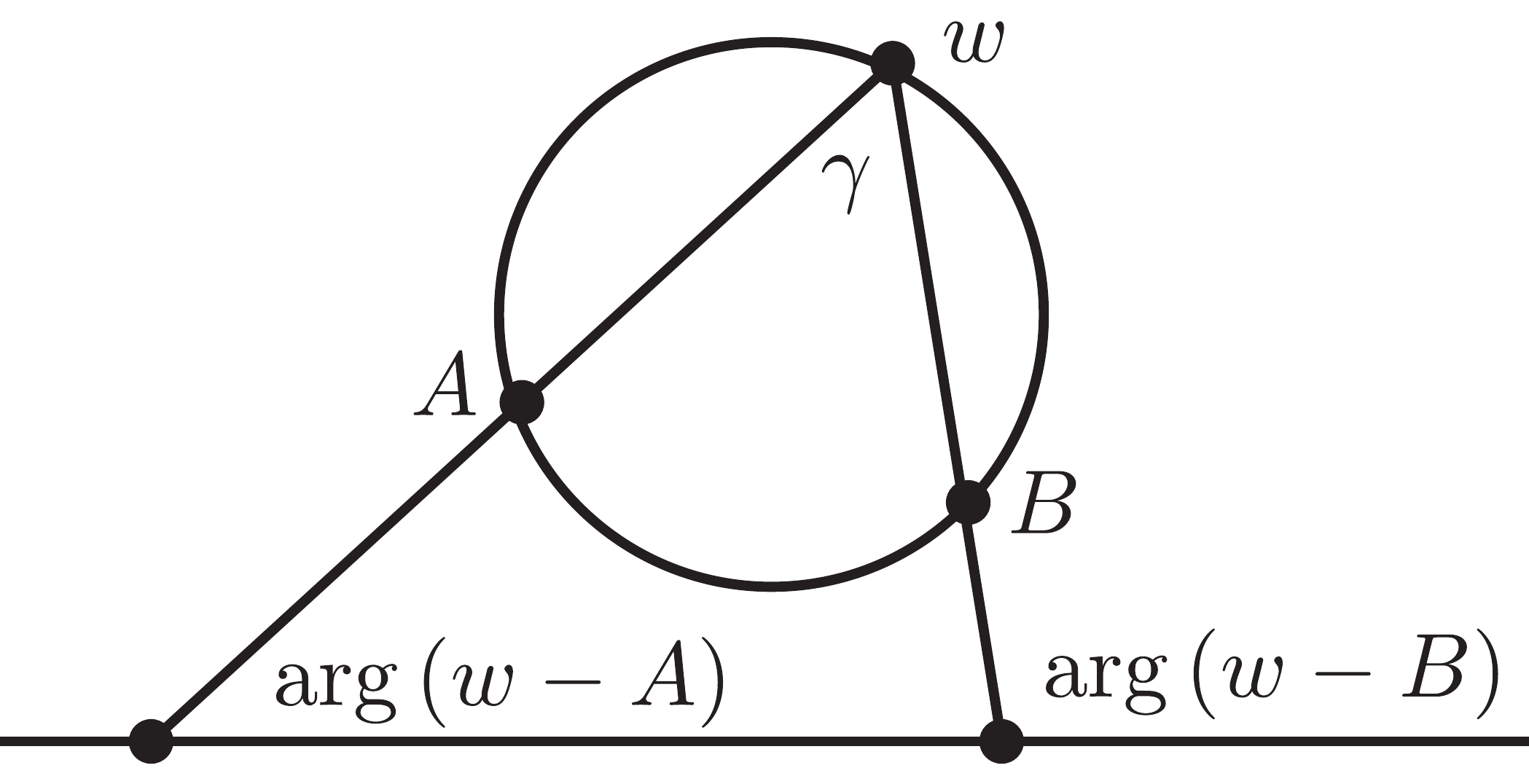}
\caption{The electric fieldlines are circular arcs given by eq. \eqref{eq:arc}, shown here for generic $A$ and $B$.}
\label{fig:arc}
\end{center}
\end{figure}

In this article, we make extensive use of the method of conformal
transformations. To obtain the director field on the sphere minimising
the free energy we use the inverse steregraphic projection, in eq.
\eqref{eq:stereo}. Another example of a conformal mapping is $\psi \left(w\right)$ in eq. \eqref{eq:map},
\begin{equation}
\label{eq:map2}
\psi \left( w\right) = \log \left( \frac{w-A}{w-B}\right),
\end{equation}
with $-A=B=id$ and $d$ real, to which we could associate electric
potential to the real part, as this holomorphic function must obey Laplace's
equation. (Note that the analogy with electrostatics made in section
\ref{subsec:electrostatic_analogy} is to aid the explanation and not
unique; we could have made an analogy with two-dimensional
fluid flow just as well.) This maps an `inhomogeneous annulus', i.e. the area bounded
by two non-concentric circles, in the complex $w$-plane to a vertical
strip in the complex $\psi$-plane. Likewise, the non-concentric
equipotential circles and electric field circles are mapped to vertical
equipotential and horizontal electric fieldlines, as in a
capacitor. It is the Mobius transformation 
\begin{equation}
\tau \left(w\right) = \frac{w+id}{w-id}
\end{equation}  
that maps the inhomogeneous annulus to a homogeneous one, \textit{i.e.} the
region bounded by two concentric circles. Consequently, this annulus
in the complex $\tau$ plane
is mapped to the vertical strip by the transformation
\begin{equation}
\psi  = \log \tau.
\end{equation}
The mapping is illustrated schematically in Fig. \ref{fig:map2}.
The equipotential (solid black) and electric field (dashed green) lines in the $w$ plane can thus be
found by setting the real and imaginairy parts of $\psi
\left(w\right)$ constant:
\begin{align}
\label{eq:real}
&\Re{\left[\psi \left( w\right)\right]} = \log \left| \frac{w-A}{w-B}\right| = \text{constant}, \\
\label{eq:im}
&\Im{\left[\psi \left( w\right)\right]} = \arg \left( \frac{w-A}{w-B}\right) = \text{constant}.
\end{align}
Eq. \eqref{eq:real} describes the circles of Apollonius characterised by inverse
points $A$ and $B$. Instead of the more familiar specification of a
circle as all the points that are a radius away from a center,
eq. \eqref{eq:real} defines a circle as the locus of points for
which the ratio of the distance to $A$ and the distance to $B$ is
constant.
It is straightforward to show that eq. \eqref{eq:real} indeed
defines circles by rewriting it into
\begin{equation}
\frac{x^2+\left(z+d\right)^2}{x^2+\left(z-d\right)^2}=C,
\end{equation}
where $C$ is a constant. Some simple algebra now leads to the usual equation of
a circle
\begin{equation}
\label{eq:circle}
x^2+\left(z-\eta\right)^2=\mathcal{R}^2.
\end{equation}
with radius, $\mathcal{R}$, and displacement, $\eta$: 
\begin{align}
\label{eq:radius}
\mathcal{R}&= \left[\left(\frac{1+C}{1-C}\right)^2-1\right]^{\frac{1}{2}}d,\\
\label{eq:displacement}
\eta &= - \frac{1+C}{1-C}\:d.
\end{align}
A look at these eqs. yields the relation between the displacements and radii of the circles:
\begin{equation}
\label{eq:desired relation}
\mathcal{R}^2=\eta^2-d^2.
\end{equation}
The electric field lines, which run perpendicular to the equipotential
lines, are also circles. Since the argument of the product of two complex numbers is
the sum of the arguments of the individual complex numbers, we can
rewrite eq. \eqref{eq:im} as
\begin{equation}
\label{eq:arc}
\arg \left( w-A\right) - \arg \left( w-B\right) = \gamma,
\end{equation}
In reference to Fig. \ref{fig:arc}, consider two fixed points $A$ and
$B$ on a circle inscribing the triangle $ABw$, where $w$ is a third point somewhere on the circular arc $AB$, then
simple geometry tells us the angle $AwB$, called $\gamma$, is
constant. Now, eq. \eqref{eq:arc} simply follows from the fact that the
sum of the angles of any (Eucledian) triangle should be $\pi$.
Alternatively, one can carry out the algebraic manipulations to find
that  
\begin{equation}
\label{eq:electric field lines}
\left(x-\epsilon\right)^2+z^2=S^2,
\end{equation} 
with radii, $S$, and displacements in the $x$-direction, $\epsilon$,
related in the following way:
\begin{equation}
S^2=\epsilon^2+d^2.
\end{equation}

\section{Experimental path through phase space}
\label{app:exppath}
In this appendix we derive the experimental trajectory through phase
space, eq. \eqref{eq:exppath}, determined by a constant $h_{min}.$ 
With a straightforward geometrical consideration we can relate
$h_{min}$ to $\Delta$:  
\begin{equation}
\label{eq:displacementvol}
\frac{\Delta}{h} = 1- \frac{h_{min}}{h}.
\end{equation}
As mentioned in the main text, the thickness is modified by a
flow of water with volume $\delta V$ that passes through the shell. We
take $\delta V > 0$ if the volume of the double emulsion droplet,
$V_{tot}$, is increased and $\delta V < 0$ if it is decreased. Upon
writing $\delta V = \frac{4}{3} \pi v^3$, we obtain for the radius of
the double emulsion droplet  
\begin{equation}
\label{eq:R}
R^3 = R'^3 + v^3.
\end{equation}
where $R'$ is the radius of the double emulsion droplet before the
flow of water. 
Since we assume that the volume of the shell, $V_{shell}$, is conserved, the inner radius changes similarly
\begin{equation}
\label{eq:a}
a^3 = a'^3 + v^3.
\end{equation}
By invoking eqs. (\ref{eq:R}) and (\ref{eq:a}) we can write for $\frac{v}{R'}$:
\begin{equation}
\label{eq:v}
\left(\frac{v}{R'}\right)^3 = \frac{\left( 1 -\frac{h}{R} \right)^3- \left(\frac{a'}{R'}\right)^3}{1-\left( 1 -\frac{h}{R} \right)^3}
\end{equation}
We wish to find the path through phase space, that is, we want to
write $\frac{\Delta}{h}$ as a function of $\frac{h}{R}$. 
We find for the displacement (eq. (\ref{eq:displacementvol}))
\begin{equation}
\label{eq:displacementvol2}
\frac{\Delta}{h} = 1- \frac{h_{min}}{R'}  \left(\frac{h}{R'}\right)^{-1}.
\end{equation}
Now by using eqs. (\ref{eq:R}) and (\ref{eq:a}), note that
\begin{equation}
\label{eq:solvethickness}
\frac{h}{R'}= \left[1+\left(\frac{v}{R'}\right)^3\right]^{\frac{1}{3}}
- \left[\left(\frac{a'}{R'}\right)^3  +\left(\frac{v}{R'}\right)^3\right]^{\frac{1}{3}}
\end{equation}
Substitution of eq. (\ref{eq:v}) into eq. (\ref{eq:solvethickness}), in turn substituted into eq. (\ref{eq:displacementvol2})  yields
\begin{equation}
\frac{\Delta}{h} = 1- \frac{u_{0}}{u}  \sqrt[3]{ \frac{1 - \left( 1- u\right)^3}{1 - \left( 1- u_0\right)^3} }
\end{equation}
where $u \equiv \frac{h}{R}$. We have chosen $\frac{a'}{R'}=1 - u_0$, with $u_0=0.0153$ the normalized thickness at which the shell becomes homogeneous. 
If we draw this trajectory in the phase diagram \ref{fig:phase_diagram}, we observe that, 
as we decrease thickness, it crosses from the bipolar regime to the non-bipolar regime via the regime of coexistence. 

\bibliography{liquid_crystals} 

\end{document}